**Unconventional dynamical covalency driven superconductivity in Nb doped $SrTiO_3$**

A. Bussmann-Holder[1], A. R. Bishop[2] and A. Simon[1]

[1]Max-Planck- Institut für Festkörperforschung, Heisenbergstr. 1, D-70569 Stuttgart, Germany

[2]Los Alamos National Laboratory, Theoretical Division, Los Alamos, NM 87545, USA

Nb doped $SrTiO_3$, the first discovered two-gap superconductor, is shown to be the most unconventional one of the known multiband superconductors, since the smaller of the two superconducting gaps follows a non BCS temperature dependence. Such a behavior stems from two cooperating effects: an extreme anisotropy in the frequency dependent interactions, involving one very soft mode and an almost vanishing interband interaction. In contrast to all other multiband superconductors, the temperature dependence of the superfluid density of Nb doped $SrTiO_3$ is predicted to exhibit an inflection point close to $T_c$ and not close to T=0.

$SrTiO_3$ is one of the best investigated perovskite oxides since it is not only of high interest from an applications point of view, but it shows also a variety of novel features upon doping and has gained substantial importance as a substrate or in layered structures. The term “quantum paraelectric” [1] was first applied to this material since pronounced momentum q=0 phonon softening is observed over a large temperature regime, which is reminiscent of a ferroelectric instability. This instability is, however, never reached since quantum fluctuations suppress the complete softening. A real ferroelectric instability occurs upon doping [2] and upon replacing $^{16}O$ by its isotope $^{18}O$ [3]. Both, doping and isotope replacement have been extensively studied and interesting new dynamics attributed to them [2, 4 – 6].

Early on it was postulated theoretically [7] that semiconductors including n-type doped $SrTiO_3$ should exhibit superconductivity since degenerate bands admit for an additional attractive inter-valley scattering channel which enhances the intra-valley attraction. Superconductivity in *n*-doped $SrTiO_3$ was discovered shortly afterwards [8 – 10] and a similar dependence of the superconducting transition $T_c$ on *n* doping was observed as more recently in cuprates and pnictides. The mechanism of superconductivity in this compound was discussed in terms of the above mentioned inter-valley scattering mechanism, but alternative approaches suggested instead that the soft mode could cause this instability [11, 12]. Since the $T_c$'s of doped $SrTiO_3$ stayed below 1K, the interest in this material diminished rapidly. Only in 1980, novel features in its superconducting properties attracted new interest in this compound, when it was found that Nb doped $SrTiO_3$ is a two-band superconductor (TBS) [13]. Even though TBS was theoretically predicted shortly after the BCS theory [14, 15], its first realization in Nb doped $SrTiO_3$ came as a surprise and remained an exception for a long time. Only after the discovery of high temperature superconductivity in cuprates [16 – 18] and $MgB_2$ [19, 20] was TBS seen to be realized in many superconductors (including the FeAs based systems [21 – 23]).

While Nb doped $SrTiO_3$ is mostly believed to be a rather conventional TBS, it is shown here that it is actually the most exceptional with respect to superconductivity since the temperature dependences of the gaps deviate strongly from what is expected for TBS. In order to demonstrate this unusual behavior the inset of Fig. 2 of Ref.13 is reproduced in Fig. 1. Here the normalized gaps are shown as a function of $T/T_c$. Typically in a TBS both normalized gaps follow the same T-dependence; this is obviously not obeyed in Nb doped $SrTiO_3$. While the larger gap follows the expected T-dependence of a TBS, the smaller gap deviates strikingly from it. Since data have been measured up to $T=0.18T_c$, the zero temperature gap values have been extracted by extrapolation which might cause uncertainties in their values. In spite of these uncertainties, there is no doubt that an anomalous behavior is exhibited by the small gap. This peculiar feature in the T-dependence of the small gap has, to our knowledge, not been addressed before.

Here we show how such unconventional effects can arise within a two band model (TBM) for superconductivity in Nb doped $SrTiO_3$.

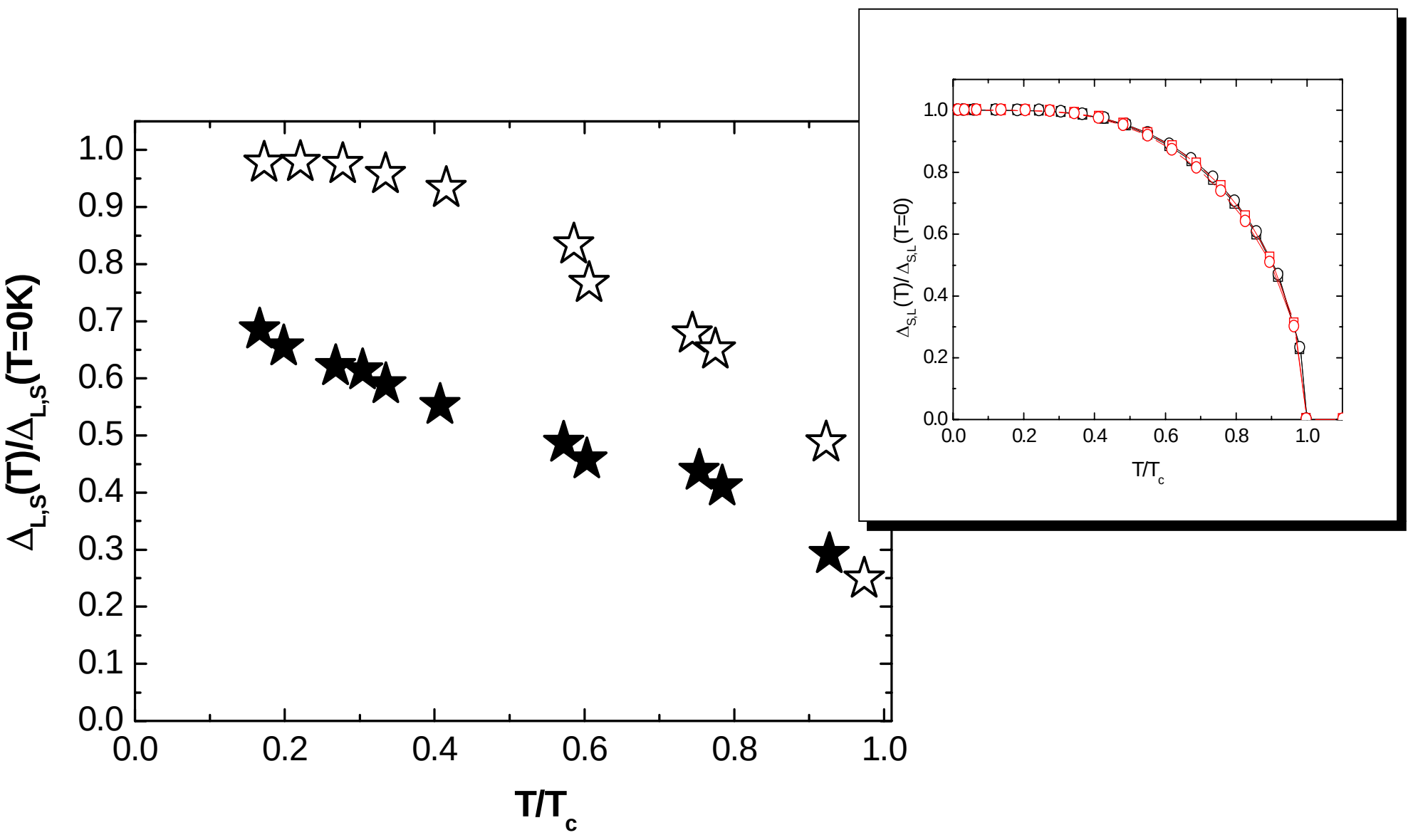

**Figure 1** (Color online) Experimentally observed dependence of the two gaps of Nb doped $SrTiO_3$ (after Ref. 13). Note, that the gaps have been measured only up to $0.18T/T_c$ leaving rather large uncertainties in the zero temperature values of the gaps, which have been used for the data normalization. The inset of Fig. 1 shows the temperature dependence of two coupled gaps for two parameter sets: in both cases $V_{1,2} > 0.02$ and $\omega_1 / \omega_2 = 20$ (red lines and symbols), $\omega_1 / \omega_2 = 0.5$ (black lines and symbols).

The TBM has been introduced in Refs. 14, 15 and further exploited in subsequent years. It has also been applied to HTSC [16- 18], $MgB_2$ [19, 20] and other superconductors, where in no case a similar T-dependence as observed for Nb doped $SrTiO_3$ has been found. The TBM starts from the assumption that two electronic bands in the vicinity of the Fermi surface contribute to the pairing mechanism with different strength. In order to guarantee a *single* $T_c$, interband interactions between both bands *must* be present, allowing pairwise exchange between both bands to take place. Principally, the involved order parameters can have different symmetries as, e.g., realized in HTSC [16 – 18]. The most frequently observed case is, however, with identical pairing symmetries in both bands. In Nb doped $SrTiO_3$ it is assumed that both order parameters have s-wave symmetry and the attractive pairing interactions are phonon mediated. These are cast in a BCS type scheme such that the Hamiltonian which is considered in the following reads:

$$H = H_0 + H_1 + H_2 + H_{12} \tag{1}$$

$$H_0 = \sum_{k_1\sigma} \xi_{k_1} c^+_{k_1\sigma} c_{k_1\sigma} + \sum_{k_2\sigma} \xi_{k_2} d^+_{k_2\sigma} d_{k_2\sigma} \tag{1a}$$

$$H_1 = -\sum_{k_1 k_1' q} V_1(k_1, k_1') c^+_{k_1+q/2\uparrow} c^+_{-k_1+q/2\downarrow} c_{-k_1'+q/2\downarrow} c_{k_1'+q/2\uparrow} \tag{1b}$$

$$H_2 = -\sum_{k_2 k_2' q} V_2(k_2, k_2') d^+_{k_2+q/2\uparrow} d^+_{-k_2+q/2\downarrow} d_{-k_2'+q/2\downarrow} d_{k_2'+q/2\uparrow} \tag{1c}$$

$$H_{12} = -\sum_{k_1 k_2 q} V_{12}(k_1, k_2) \{ c^+_{k_1+q/2\uparrow} c^+_{-k_1+q/2\downarrow} d_{-k_2+q/2\downarrow} d_{k_2+q/2\uparrow} + h.c. \}. \tag{1d}$$

Here $\xi_{k_i}$ are the momentum *k* dependent energies in band *i*, with creation and annihilation operators $c^+, c,\ d^+, d$ ; $V_i, V_{12}, V_{21}$ are the effective attractive interactions in band *i*=1, 2, and the interband interactions which mediate pairwise exchange between the two bands. In Ref. 13 it was suggested that the considered bands can be related to two sheets centered at $k = 0$, where one refers to the lowest anisotropic conduction band while the second band is isotropic and upward shifted by 20meV. Instead we identify these bands as arising from the doped Ti $d^1$ states and in-gap states (IGS) caused by strong p-d hybridization as recently inferred from photoemission spectroscopy [24]. This implies that the $d^1$ band is the primary cause of superconductivity, whereas the IGS related band shows *induced* superconductivity due to the strong p-d hybridization. It is important to note, that the IGS band extends over a broad energy range below the Ti $d^1$ coherent states (see Fig. 5 of Ref. 24) and overlaps smoothly with these thus admitting for interband interactions.

From Eqs. 1 the gap equations are readily obtained by standard techniques and are explicitly given by:

$$< c^+_{k_1\uparrow} c^+_{-k_1\downarrow} > = \frac{\Delta_{k_1}}{2E_{k_1}} \tanh\frac{\beta E_{k_1}}{2} = \Delta_{k_1} \Phi_{k_1} \tag{2a}$$

$$< d^+_{k_2\uparrow} d^+_{-k_2\downarrow} >= \frac{\Delta_{k_2}}{2E_{k_2}} \tanh\frac{\beta E_{k_2}}{2} = \Delta_{k_2}\Phi_{k_2} \tag{2b}$$

$$\Delta_{k_1} = \sum_{k_1'} V_1(k_1,k_1')\Delta_{k_1'}\Phi_{k_1'} + \sum_{k_1} V_{1,2}(k_1,k_2)\Delta_{k_2}\Phi_{k_2} \tag{2c}$$

$$\Delta_{k_2} = \sum_{k_2'} V_2(k_2,k_2')\Delta_{k_2'}\Phi_{k_2'} + \sum_{k_1} V_{2,1}(k_2,k_1)\Delta_{k_1}\Phi_{k_1} \tag{2d}$$

From these equations coupled gaps are obtained which have to be evaluated simultaneously and selfconsistently in order to calculate their temperature dependence and $T_c$. Obviously, a number of parameters enter the problem: the four pairing interactions, and the frequency dependencies of these interactions. In order to minimize the parameter space, we take $V_{1,2} = V_{2,1}$. The number of parameters can be further reduced by requiring that the large and small gaps have the experimentally observed zero temperature values, namely $\Delta_L \cong 0.09 meV, \Delta_S \cong 0.06 meV$ and $T_c \approx 0.2$-$0.4$K. With this choice only three parameters remain, which are used as variables to find the origin of the anomaly observed for the small gap. These are the frequency ranges for the two intraband interactions and the magnitude of the interband interaction. For any given values of these quantities there exists *only one set* of intraband interactions $V_1, V_2$ for which the gap values exhibit the correct magnitudes. If the momentum sums in the coupled gaps are of the same order of magnitude or deviate from each other (independent of whether the large or small gap is concerned) by as much as 20% and $V_{1,2}$ is larger than 0.02, both gaps obey the BCS temperature dependence (see inset to Fig. 1) without showing any anomalies. It is, however, important to mention, that in spite of the fact that the gap values are fixed, $T_c$ shows small variations of the order 0.02K upon changing either the k-space integration or the interband interaction.

A very different behavior in the T-dependence of $\Delta_S, \Delta_L$ occurs when the integration limits are varied beyond 20% and the interband interaction $V_{1,2} < 0.02$. This is shown in Fig. 2a where for all values of $\omega_1 / \omega_2$ ($\omega_1, \omega_2$ are the momentum integrations) $V_{1,2}$ is kept constant: $V_{1,2} = 0.015$.

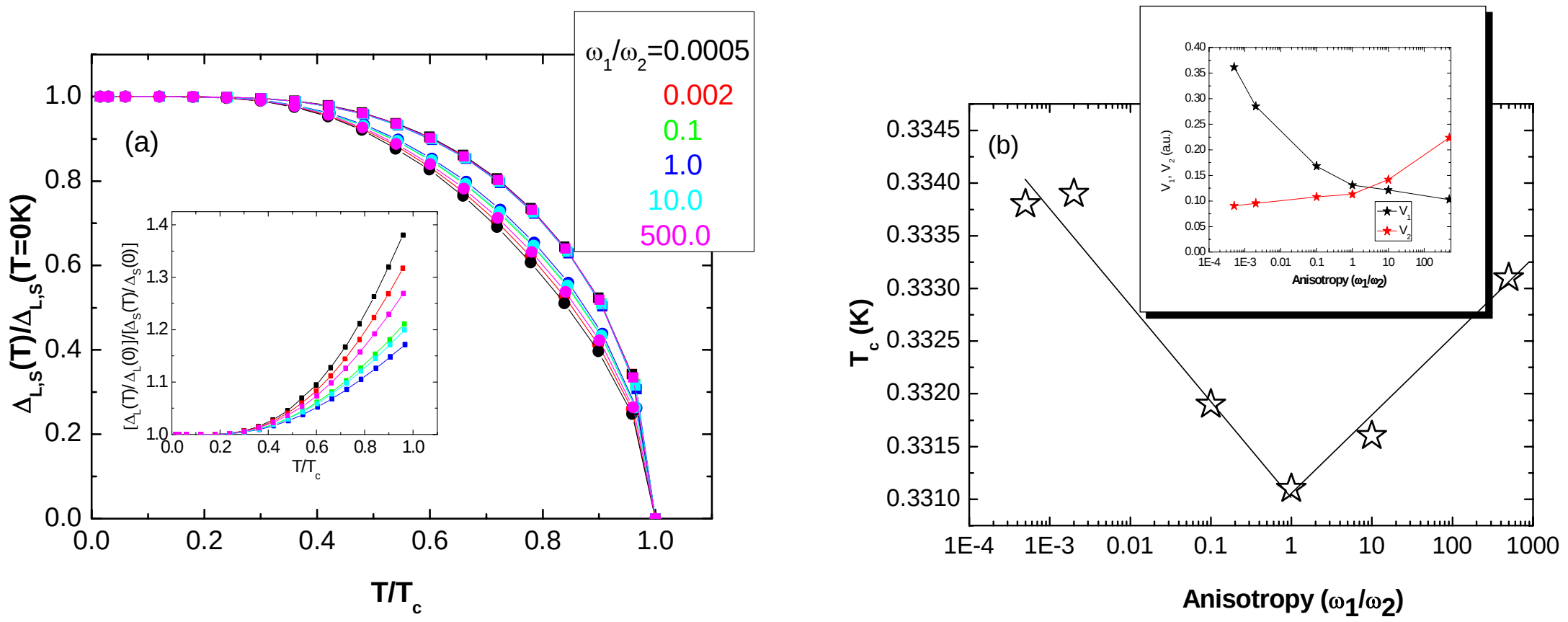

**Figure 2** (Color online) (a) Normalized large and small gaps $\Delta_L, \Delta_S$ versus T/T$_c$ for various values of $\omega_1 / \omega_2$, as shown in the color code depicted in the figure. The inset shows the ratio of the normalized gaps (gap anisotropy) as a function of T/T$_c$. (b) The dependence of T$_c$ on the anisotropy $\omega_1 / \omega_2$. The inset shows the selfconsistently determined values of the intraband coupling constants $V_1, V_2$ on the same anisotropy.

With increasing and decreasing values of the ratio $\omega_1 / \omega_2$, the smaller gap starts to deviate from the conventional behavior and adopts an anomalous dependence between 0.03<T/T$_c$<0.95. This anomaly is larger when $\omega_1 / \omega_2$ is very small as compared to the reversed case. This is shown in the inset to the Fig. 2a where the normalized gap anisotropy is depicted and the magenta and red curves correspond to the reversal of the leading frequency. Obviously, the most anomalous behavior results from the very small frequency range for the large gap which suggests that here the soft mode plays a decisive role. Simultaneously, it is necessary that a large momentum cutoff is involved in the small gap integration. Another consequence of the variation in the momentum cutoff is a systematic variation in $T_c$ (see Fig. 2b). With decreasing and increasing $\omega_1 / \omega_2$ $T_c$ increases with steeper increase for the former case. It is, nevertheless, obvious from the figure that even the most anisotropic case, i.e., $\omega_1 / \omega_2 = 0.0005$ does not fully reproduce the observed anomaly. In order to clarify this point further, this ratio has been kept constant ($\omega_1 / \omega_2 = 0.001$) and the interband interaction varied. With decreasing $V_{1,2}$ the smaller gap is depressed at temperatures $T/T_c>0.4$ (see Fig. 3) and shows a pseudo superconducting instability for the smallest $V_{1,2} = 0.0005$ around 0.275K where it drops substantially. This small value of the interband coupling corresponds to the almost completely decoupled case where for a two band superconductor without gap coupling two successive superconducting transitions are expected corresponding to phase separation. Seemingly, the system tends to this case but due to the finite coupling $\Delta_S$ closes at the larger $T_c$. Another effect of decreasing the interband coupling is an increase in $T_c$ which amounts to 10% for the smallest coupling as compared to the largest coupling. Typically the interband interaction causes enormous increases in $T_c$ with increasing values of $V_{1,2}$ as has been shown for the case of Al doped $MgB_2$ [25] and cuprates [26]. Here, however, the opposite case is observed which is also caused by the condition that the zero temperature gap values remain unchanged. The reversed trend in $T_c$ cannot be explained only by the constancy in the gap values but is also a consequence of the large anisotropy in the frequency cutoffs. The small value of $\omega_1 / \omega_2$ used here implies that the effective coupling constant $V_1$ of the large gap is substantially larger than that of the small gap, $V_2$, and increases with either increase in anisotropy or decreasing interband coupling thus leading to the enhancement of $T_c$. The dependence of the intraband couplings on either of these quantities is shown as inset to Fig. 2b and as inset to Fig. 3, respectively.

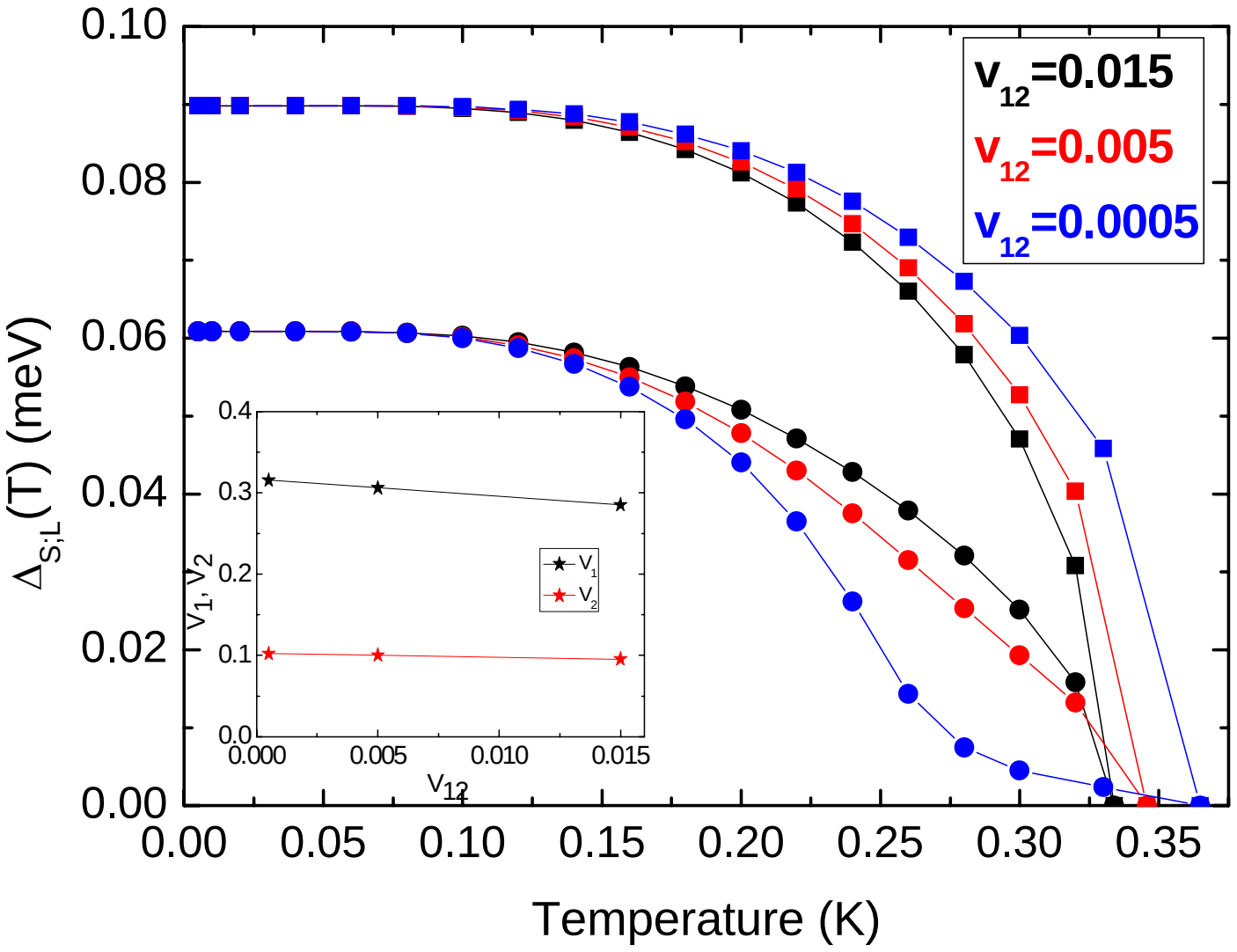

**Figure 3** (Color online) $\Delta_{S,L}$ as a function of temperature T for varying interband coupling constant $V_{12}$ with values as given in the figure. The inset shows the selfconsistently determined intraband coupling constants $V_1, V_2$ as a function of the same quantity.

Clearly the variations in the cutoff frequency ratio as well as that of the interband interaction $V_{12}$ determine these quantities and it is seen that $V_1$ is always substantially larger than $V_2$ for decreasing $V_{12}$, and as long as $\omega_1 / \omega_2 < 3$, where a reversal in the leading coupling takes place.

An important predicted consequence of the anomalous T-dependence of the smaller gap is related to the T-dependence of the superfluid density $\rho_s(T)$ which can be calculated by standard methods from the gaps. For cuprates and $MgB_2$ is has been shown [16 – 20] that the superfluid density exhibits an inflection point close to T=0K in its T-dependence when two gaps with very different values are present. In the case of Nb doped $SrTiO_3$ the gaps do not differ very much from each other ( $\Delta_L / \Delta_S \approx 1.5$ , for comparison $\Delta_L / \Delta_S \approx 4.5$ in $MgB_2$) and a suppression of the low temperature inflection point is expected. Instead, an inflection point appears close to $T_c$ (see Fig. 4) which is absent if both gaps follow a BCS type T-dependence (as, e.g., shown in the inset to Fig. 1 and black symbols and lines in Fig. 4). In order to emphasize this anomaly more clearly the derivatives of the superfluid densities have been calculated and are shown in the upper inset to Fig. 4. At $T/T_c>0.6$ a non-monotonic behavior sets in, strongly contrasting with the case where a BCS temperature dependence of the gaps is present (black lines and symbols in Fig. 4 and results shown in the inset to Fig. 1). The individual components from the two gaps contributing to the superfluid density are shown in the lower inset to Fig. 4. It is important to mention here that this behavior is independent of the percentage admixture of the two gaps contributing to the superfluid density.

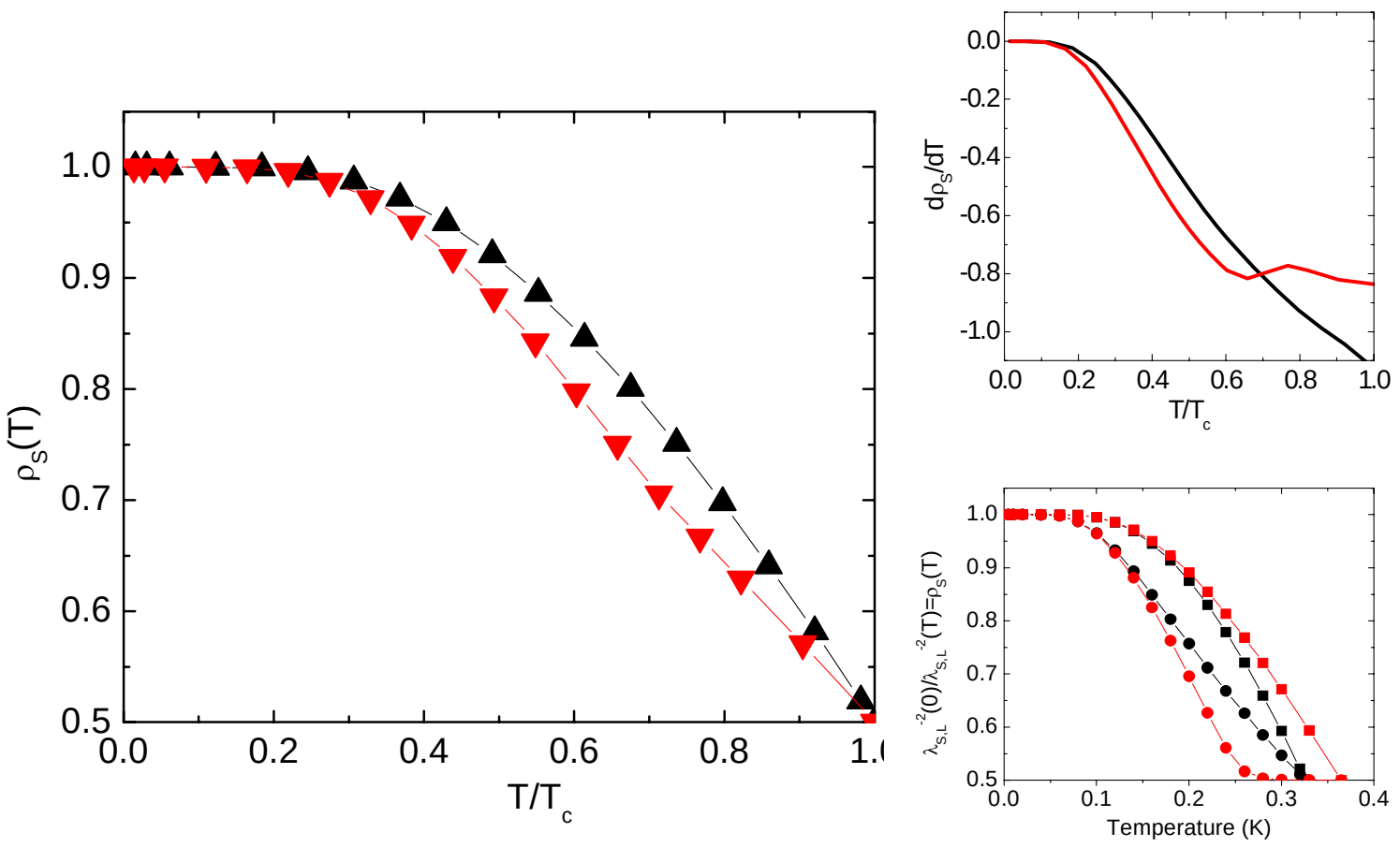

**Figure 4** (Color online) Superfluid density $\rho_s(T)$ as a function of T/T$_c$. The black line and symbols refer to the typical BCS temperature dependence. The red line and symbols are calculated using the results shown in Fig. 3 (blue lines and symbols). The upper inset is the temperature derivative of the results in the main figure. The lower inset shows the individual component to $\rho_s(T)$.

Regarding the experimental data for the TBS Nb doped $SrTiO_3$ (see Fig. 1), the results presented above show that the anomalous temperature dependence of the smaller gap stems from two effects: i) a very small interband coupling $V_{12}$; and ii) a large anisotropy in the frequency integration where a small value of $\omega_1 / \omega_2$ is more favorable than a large one. The small interband coupling shows that the two-gap features that have only been observed in the Nb doped compound must be a consequence of the formation of p-d hybridized IGS which are not observed in oxygen vacancy doped $SrTiO_3$. The closeness of these states to the Ti d$^1$ band facilitates weak interband interactions which in part cause the deviations of the small gap from BCS like behavior. An additional prerequisite is the frequency anisotropy which suggests that the soft mode of $SrTiO_3$ that persists with Nb doping is strong coupling and the cause of superconductivity as already outlined early on [11, 12]. Besides causing superconductivity, it is also fundamental to the temperature anomaly of the small gap. Regarding the Nb-doping dependence of $T_c$ as reported in Ref. 13, lower doping levels as considered here (namely optimum doping) reduce the Ti $d^1$ density of states and thereby diminish $V_1$ which naturally reduces $T_c$. In addition, also the interaction between the IGS and these states is diminished which for zero doping leads to a collapse of superconductivity and a metal insulator transition. Higher than optimum doping levels, on the other hand, split the Ti $d^1$ states apart from the IGS, i.e., the interband interaction vanishes and and a single gap system emerges with naturally lower values of $T_c$ than observed in a TBM.

In conclusion, it has been shown that TBS in Nb doped $SrTiO_3$ is very different from TBS in cuprate HTS, $MgB_2$, FeAs based compounds and other known two band superconductors. Important ingredients for the observed deviations in the T-dependence of the small gap are a strong coupling to the soft mode and a small interaction between the two involved bands. Since the oxygen isotope-replaced $SrTi^{18}O_3$ system exhibits a ferroelectric instability, we predict that Nb doping of this compound will not induce TBS or superconductivity at all. The quantum paraelectric state where the soft mode remains finite at very low frequency and low temperatures together with the p-d hybridized IGS are the origin of TBS. The related superfluid density is predicted to show an inflection point in the vicinity of $T_c$ in contrast to the above mentioned TBS where the inflection point appears close to T=0K.

**Acknowledgement** It is a pleasure to acknowledge helpful discussions with R. K. Kremer.

**References**

1. K. A. Müller and H. Burkard, Phys. Rev. B **19**, 3593 (1979).
2. J. G. Bednorz and K. A. Müller, Phys. Rev. Lett. **52**, 2289 (1984).
3. M. Itoh, R. Wang, Y. Inaguma, T. Yamaguchi, Y. J. Shan, and T. Nakamura, Phys. Rev. Lett. **82**, 3540 (1999).
4. A. Bussmann-Holder, H. Büttner, and A. R. Bishop, J. Phys.: Cond. Mat. **12**, L115 (2000).
5. A. Bussmann-Holder and A. R. Bishop, Europhys. Lett. **76**, 945 (2006).
6. A. Bussmann-Holder, H. Büttner, and A. R. Bishop, Phys. Rev. Lett. **99**, 167603 (2007).
7. M. L. Cohen, Phys. Rev. **134**, A511 (1964).
8. J. F. Schooley, W. R. Hosler, and M. L. Cohen, Phys. Rev. Lett. **12**, 474 (1964).
9. C. S. Koonce, M. L. Cohen, J. F. Schooley, W. R. Hosler, and E. R. Pfeiffer, Phys. Rev. **163**, 380 (1967).
10. N. E. Phillips, J. C. Ho, D. P. Woody, J. K. Hulm, and C. K. Jones, Phys. Lett. **29A**, 356 (1969).
11. J. Appel, Phys. Rev. Lett. **17**, 1045 (1966).
12. J. Appel, Phys. Rev. **180**, 508 (1969).

13. G. Binnig, A. Baratoff, H. E. Hoenig, and J. G. Bednorz, Phys. Rev. Lett. **45**, 1352 (1980).
14. H. Suhl, B. T. Matthias, and L. R. Walker, Phys. Rev. Lett. **3**, 552 (1959).
15. V. A. Moskalenko, Fiz. Metal Metalloved. **8**, 503 (1959).
16. R. Khasanov, A. Shengelaya, A. Maisuradze, F. La Mattina, A. Bussmann-Holder, H. Keller, and K. A. Müller, Phys. Rev. Lett. **98**, 057007 (2007).
17. R. Khasanov, S. Strässle, D. DiCastro, T. Masui, S. Miyasaka, S. Tajima, A. Bussmann-Holder, and H. Keller, Phys. Rev. Lett. **99**, 237601 (2007).
18. R. Khasanov, A. Shengelaya, J. Karpinski, A. Bussmann-Holder, H. Keller, and K. A. Müller, J. Supercond. and Novel Magnetism **21**, 1557 (2008).
19. Amy Y. Liu, I. I. Mazin, and J. Kortus, Phys. Rev. Lett. **87**, 087005 (2001).
20. E. J. Nicol and J. P. Carbotte, Phys. Rev. B **71**, 054501 (2005).
21. Cong Ren, Zhao-Sheng Wang, Hui-Qian Luo, Huan Yang, Lei Shan, and Hai-Hu Wen, Phys. Rev. Lett. **101**, 257006 (2008).
22. R. Khasanov, D. V. Evtushinsky, A. Amato, H.-H. Klauss, H. Luetkens, Ch. Niedermayer, B. Büchner, G. L. Sun, C. T. Lin, J. T. Park, D. S. Inosov, and V. Hinkov, Phys. Rev. Lett. **102**, 187005 (2009).
23. R. Khasanov, K. Conder, E. Pomjakushina, A. Amato, C. Baines, Z. Bukowski, J. Karpinski, S. Katrych, H.-H. Klauss, H. Luetkens, A. Shengelaya, and N. D. Zhigadlo, Phys. Rev. B **78**, 220510(R) (2008).
24. Y. Ishida, R. Eguchi, M. Matsunami, K. Horiba, M. Taguchi, A. Chainani, Y. Senka, H. Ohashi, H. Ohta, and S. Shin, Phys. Rev. Lett. **100**, 056401 (2008).
25. A. Bussmann-Holder and A. Bianconi, Phys. Rev. B **67**, 132503 (2003).

26. R. Micnas, S. Robaskiewicz, and A. Bussmann-Holder, in “Superconductivity in Complex Systems”, Springer Series Structure and Bonding, Ed. K. A. Müller and A. Bussmann-Holder (Springer, Heidelberg), **114** (2005).